# A theoretical analysis and determination of the technical requirements for a Bragg diffraction-based cold atom interferometry gravimeter

Hu Qingqing[1,2], Yang Jun[1,2], Luo Yukun[1,2], Jia Aiai[1,2], Wei Chunhua[1,2], Li Zehuan[1,2]

(1.College of Mechatronics Engineering and Automation, National University of Defense Technology, Changsha 410073, China;

2. Interdisciplinary Center for Quantum Information, National University of Defense Technology, Changsha 410073, China)

**Abstract:** We present here a new type of cold atom interferometry gravimeter based on Bragg diffraction, which is able to increase the gravity measurement sensitivity and stability of common Raman atom gravimeters significantly. By comparing with Raman transition, the principles and advantages of Bragg diffraction-based atom gravimeters have been introduced. The theoretical model for a time-domain Bragg atom gravimeter with atomic incident direction parallels to the wave vector of Bragg lasers has been constructed. Some key technical requirements for an $n$th-order Bragg diffraction-based atom gravimeter have been deduced, including the temperature of atom cloud, the diameter, curvature radius, frequency, intensity, and timing sequence of Bragg lasers, etc. The analysis results were verified by the existing experimental data in discussion. The present study provides a good reference for the understanding and construction of a Bragg atom gravimeter.

**Keywords:** cold atom interferometry, gravimeter, $n$th order Bragg diffraction, large momentum transfer

## 1 Introduction

Since the first cold atom interferometer achieved by Steven Chu's team in 1991[1], atom interferometers have been demonstrated remarkable prospects in inertial sensing and fundamental physics due to their high measurement accuracy and sensitivity. Applications stretch from high-precision gyroscopes[2], gravimeters[3], gravity gradiometers[4], to the measurements of fine structure constant[5] and gravitational constant[6]. Furthermore, atom interferometry gravimeter based on stimulated Raman transition are on the stage of commercialization [7].

From the sensitivity formula of atom gravimeter $S=\mathbf{k}_{eff}\mathbf{g}T^2$ ($\mathbf{k}_{eff}$ is the effective wave vector of Raman laser, which decides the momentum separation; $T$ is the interval between two laser pulses, representing atomic free evolution time), we see two methods to increase its sensitivity. One is increasing the free evolution time of atom, therefore atom interferometer long up to 100m[8] has been realized. This scheme obviously doesn't suit for practical application except fundamental physics research; the other method is increasing the effective laser wave vector, such as sequential two-photon Raman transition[9], $n$-order Bragg diffraction[10], Bloch oscillations in optical lattices[11] et al large momentum transfer (LMT) scheme. Due to its unique advantages, $n$-order Bragg diffraction becomes one of the most promising candidate technology among all the LMT schemes.

Although atom interferometers based on Bragg diffraction have been reported already[12,13], and gravity measurement based on Bragg diffraction $^{87}$Rb atoms has been realized by Debs et al[13], the theoretical research of Bragg atom gravimeter is few, especially for laser's incident direction parallels atomic free fall direction. Giltner et al[14] deduced the analytic expression for Bragg diffraction probability for the laser standing perpendicular to the atomic free fall direction; Blakie et al[15] analyzed the Bragg momentum spectrum and resonance conditions for the two energy states BEC atoms using the mean-field approximation method; Müller et al[16] analyzed the effect of laser intensity, interaction time, and pulse shape on the effective Rabi frequency, diffraction loss and phase shift in case of quasi-Bragg diffraction; Szigeti et al[17] simulated the effect of atomic momentum broadening on the fringes contrast. Most of these theoretical models suit only for space-domain atom interferometer with laser standing waves (SWs) perpendicular to atomic incident direction. In this paper, we present a time-domain Bragg atom gravimeter with SWs parallel atomic incident direction. We firstly introduce the basic principles, then derive the technical requirements in detail using the constructed physical model, and finally we compare our results with the existing experimental data.

## 2 Basic principles

### 2.1 First order Bragg diffraction

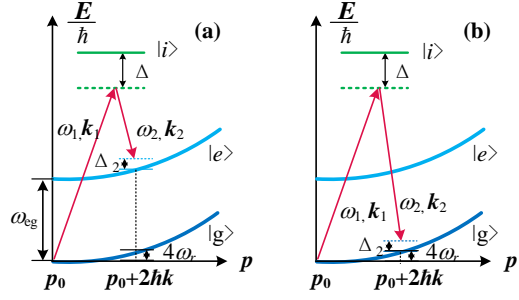

Fig.1. Comparison between (a) two-photon Raman transition and (b) first order Bragg diffraction.

Figure 1 shows a comparison between first order Bragg diffraction and two-photon Raman transition of $^{87}$Rb atom. In Fig. 1, $|g\rangle$ and $|e\rangle$ represent atomic two hyperfine ground states with an energy interval of $\omega_{eg} \approx 6.8$ GHz, $|i\rangle$ represents the excited state, $\Delta$ and $\Delta_2$ were single-photon and two-photon detunings. Two-photon Raman transition process can be described as follows: a pair of counter-propagating Raman lasers irradiate atoms and transit them from $|g, p\rangle$ to $|e, p + \hbar k_{eff}\rangle$ by stimulated absorb a photon of frequency $\omega_1$ and reverse stimulated emission a photon of frequency $\omega_2$ via the intermediate virtual state $|i\rangle$. Two-photon detuning equals

$$\Delta_{2R} = (\omega_1 - \omega_2)_R - (\omega_{eg} + 4\omega_r + 2\mathbf{k} \cdot \mathbf{v}) \tag{1}$$

where $\omega_r = \hbar k^2/2M \approx 2\pi \times 3.77$ kHz is photon recoil frequency shift, $M$ is $^{87}$Rb atomic mass, and $2\mathbf{k} \cdot \mathbf{v}$ is Doppler shift.

Comparing Fig. 1 (a) and (b) we find two main differences between the first-order Bragg diffraction and two-photon Raman transition. Firstly, atoms stay in the same lower energy state $|g\rangle$ both before and after the Bragg diffraction; secondly, two-photon detuning in Bragg diffraction equals

$$\Delta_{2B} = (\omega_1 - \omega_2)_B - (4\omega_r + 2\mathbf{k} \cdot \mathbf{v}) \tag{2}$$

Two-photon detuning equals approximately zero in most cases, therefore the frequency difference of the two counter-propagating lasers in Bragg diffraction is about 6.8GHz smaller than that of Raman transition.

These two differences make Bragg atom gravimeter have unique advantages as follows: **a).** the Bragg transition frequency is typically several orders of magnitude less than for a Raman transition (~1MHz compared to 10 GHz), which simplifies the optical and electronic systems required to effect beam splitting; **b).** atoms always remain in the lower energy state, so spontaneous emission, which destroys the coherence of the atom beam, is not an issue during the free flight of the beams therefore interferometers with long arm lengths are allowed; **c).** atoms are in the same state during the two paths of the interferometer, thus the atomic phase is not affected by temporal fluctuations in the SWs and is less sensitive to external fields[12]; **d).** the same atom state before and after Bragg diffraction makes $n$-order LMT possible.

## 2.2 $n$-order Bragg diffraction

Partial transition diagram for atomic $n$th-order Bragg diffraction is shown in Fig. 2, which can be viewed as a $2n$-photon stimulated Raman process. The atoms start from a state with momentum $p_z = 0$, stimulated absorb and reverse emission $n$ photons of frequency $\omega_1$ and $\omega_2$ respectively, end in the state $|g, 2n\hbar k\rangle$ via the $2n$-1 intermediate virtual states $|i, \hbar k\rangle$, $|g, 2\hbar k\rangle$, …, $|i, (2n-1)\hbar k\rangle$. The parabolas correspond to kinetic energy $p^2/2M$, $\Delta_i$ ($i$=1, 2 … $2n$-1) denotes the detuning of laser frequency from the $i$th resonance transition, $\delta_n$ represents the resonant condition for $n$th-order Bragg diffraction.

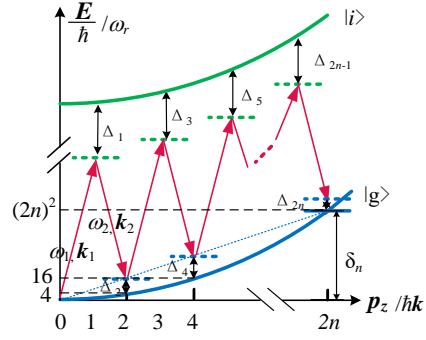

Fig.2. Partial transition diagram for atomic *n*th order Bragg diffraction.

## 2.3 Atom interferometry gravimeter based on *n*-order Bragg diffraction

Figure 3 shows the basic operating principle for the *n*th-order Bragg diffraction-based atom interferometer in the Mach-Zehnder configuration. The incoming atom cloud is coherently split into two parts by the first π/2 pulse, which functions as a beam splitter and causes half of the incoming atoms experience a momentum transfer of $2n\hbar k$. The two atom clouds then propagate freely for an "interrogation time" *T* along two different paths. A second π pulse, just like a mirror, is then applied to exchange the momentum states of the two atom clouds. After another "interrogation time" *T*, the two atom clouds are recombined coherently by the third π/2 pulse. During this process, each of the three Bragg pulses imprints a phase $\phi_j$ ($\phi'_j$) (where *j*=1, 2, 3, corresponding to the 1st, 2nd and 3rd optical pulse respectively) onto the atomic wavepackets. At last, the two atomic wavepackets accumulate a phase difference $\Delta\phi$. For uniform gravity acceleration, $\Delta\phi$ calculates to

$$\Delta\phi = n\left(2\mathbf{k}\cdot\mathbf{g}T^2 - 2\pi\alpha T^2\right) \tag{3}$$

where *n* is Bragg diffraction order, and $\alpha \approx 25.1\text{MHz/s}$ is the chirping rate of the laser frequency to compensate Doppler shift caused by gravity. Eq.(3) shows that Bragg diffraction could increase the phase difference by *n* times compared with the Raman atom gravimeter.

To achieve the interference signal, we need to measure the relative atom population $P_1$ ($P_2$) with momentum $\mathbf{p}$ ($\mathbf{p}+2n\hbar\mathbf{k}$) in Output channel 1(2) (see Fig. 3) as

$$\begin{aligned} P_1 &= 1/2\left(1+V\cos\Delta\phi\right) \\ P_2 &= 1/2\left(1-V\cos\Delta\phi\right) \end{aligned} \tag{4}$$

where $V \leq 1$ is the contrast of the interference fringes.

Before scanning the phase difference $\Delta\phi$ to produce a sinusoidal interference curve, we need to find the resonant chirping rate $\alpha_0$ first, which keeps $2\mathbf{k}\cdot\mathbf{g}-2\pi\alpha=0$ for all *T* and can be identified by linear sweeping the frequency chirping rate α under different *T* to make multiple interference fringes[18], as shown in Fig. 4 (black dashed line, red dotted line, and blue solid curves respectively corresponding to *T*=40, 50, 60ms, *V*=1, *n*=1). Consequently, gravity can be calculated as

$$\mathbf{g} = \frac{\pi\alpha_0}{\mathbf{k}} \tag{5}$$

Afterwards, a more precision value of local gravity can be deduced from the sinusoidal interference curve produced by scanning the phase of the third Bragg laser using Eqs.(3) and (4), with the chirping rate set in the resonant chirping rate $\alpha_0$.

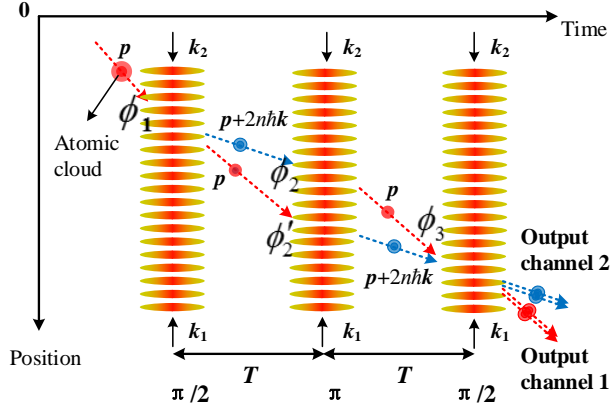

Fig. 3. Space-time diagram of the *n*th order Bragg diffraction-based atom gravimeter.

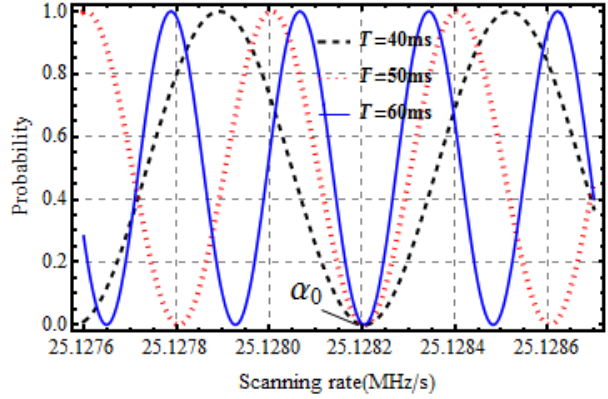

Fig. 4. Identifying of resonant chirping rate by making multiple interference fringes of different *T*.

## 3 Technical requirements

### 3.1 Longitudinal temperature of atom cloud

When Bragg pulses irradiate atoms, the finite duration $\tau$ gives a specific spread of Fourier frequency (approximately $\tau^{-1}$). If the frequency spread is too large, the atoms would be transited to other unwanted momentum states; whereas if it is too narrow, only small portion of atoms at the center momentum state would be transferred, both led to low diffraction efficiency and fringe contrast. Therefore, Bragg pulse frequency broadening should be much smaller than the 1st-order Bragg diffraction resonance bandwidth $\delta_B$ ($4\omega_r=2\hbar k^2/M=2\pi\times15.1$ kHz) to avoid spurious diffraction, and much larger than the atomic non-zero momentum width to improve diffraction efficiency

$$\frac{2k\Delta p_\parallel}{M} \ll \frac{1}{\tau} \ll \delta_B \tag{6}$$

Satisfying these two conditions simultaneously is essential for complete Bragg diffraction to occur. Whilst we cannot change the Bragg condition for $^{87}$Rb, we can change the length of our pulses, their shape in Fourier space, and the momentum width of the atom cloud. To make the above equation solvable, requiring atomic longitudinal momentum broadening $\Delta p_\parallel \ll \hbar k$, and the corresponding longitudinal temperature $T_\parallel \ll 0.36 \mu k$.

### 3.2 Diameter and curvature radius of Bragg laser

The movement of cold atoms in the three Bragg pulses is shown in Fig. 5. When $t=0$, assuming the atom cloud have a Gaussian density distribution with transverse radius $r_0$ (density reduced to $1/e$ of the maximum), then the transverse radius of atom cloud at time $t$ is

$$r^2(t) = r_0^2 + v_\perp^2 t^2 \tag{7}$$

where $v_\perp = \sqrt{k_B T_\perp/M}$ is atomic transverse velocity, and $T_\perp$ is the equivalent transverse temperature. To ensure Bragg lasers irradiate all atoms, the $1/e^2$ diameter should be larger than the transverse diameter of atom cloud

$$w > 2\sqrt{r_0^2 + v_\perp^2 (t_0 + 2T)^2} \quad (8)$$

where $t_0$ and $T$ represent the moment when the first $\pi/2$ pulse acts on atoms and the interval between two pulses.

Since each of the three Bragg pulses imprints its phase $\phi_j$ on atomic phase, the curved wavefront of laser pulses would cause a phase error on the interference signal. Thus a larger curvature radius of Bragg laser is required so that the atoms approximately feel a plane wavefront. In experiments, Bragg laser always owns a parabolic wavefront, as shown in Fig. 5, thus the total phase error caused by wavefront curvature of the three Bragg lasers is

$$\delta\phi_{laser} = \delta\phi_1 - 2\delta\phi_2 + \delta\phi_3 = \frac{k_{eff}}{R(z)} v_\perp^2 T^2 \quad (9)$$

where $R(z)$ is the curvature of Bragg laser in the position where atom was irradiated. Combing Eq.(3), we know that in order to obtain sub-µGal ($10^{-9}$g) level gravity measurement accuracy, requiring wavefront curvature of Bragg laser satisfies

$$R(z) \geq \frac{v_\perp^2}{1\mu\text{Gal}} \quad (10)$$

Taking typical parameters [19] in Fig. 5 with: $r_0$=1.5mm, $t_0$=20ms, $T$=50ms, $T_\perp$=5µk (atoms with longitudinal velocity selection), and $T_\perp$=0.36µk (BEC atoms). Substituting these parameters into Eqs.(8) and (10), we obtain the corresponding Bragg laser's $1/e^2$ diameter $w$>6mm or 1.7mm, curvature radius $R(z)$>48.8km or 3.5km.

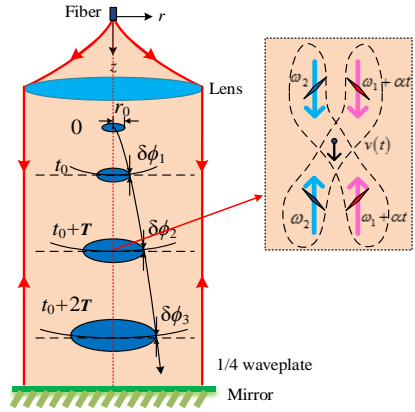

Fig. 5. Diagram of Bragg lasers interacting with atoms.

### 3.3 Frequency difference and detuning of Bragg laser

According to the $n$-order Bragg diffraction theory described in section 1.2, and the momentum-energy relation of $^{87}$Rb hyperfine ground state shown in Fig. 2, the detuning between laser frequency and transition frequency of the intermediate momentum state is

$$\Delta_m = \begin{cases} \Delta + \left(m^2\omega_r - \frac{m-1}{2}\omega_{eff}\right), & m=1,3\cdots 2n-1 \\ \frac{m}{2}\omega_{eff} - m^2\omega_r, & m=2,4\cdots 2n \end{cases} \quad (11)$$

where $\omega_{eff} = \omega_1 - \omega_2$ is frequency difference of the two Bragg pulses. The detuning $\Delta_m = 0$ when $m=2n$, thus

$$\omega_{eff} = 4n\omega_r = n\delta_B \quad (12)$$

To compensate Doppler shift caused by gravity, the laser frequency is linear chirped at rate of $\alpha \approx 25.1\text{MHz/s}$. Then the resonance conditions for $n$-order Bragg diffraction changes to

$$\delta_n(t) = n\delta_B + 2kgt \quad (13)$$

Using Eq.(12), the detuning to intermediate state can be simplified to

$$\Delta_m = \begin{cases} \Delta + \left[m^2 - 2n(m-1)\right]\omega_r, & m=1,3\cdots 2n-1 \\ m(2n-m)\omega_r, & m=2,4\cdots 2n \end{cases} \quad (14)$$

Under large detuning condition ($\Delta \gg \omega_r$), the probability of atoms transited to excited state can be estimated as

$$P_i(t) \approx \left(\frac{\Omega_0}{2\Delta}\right)^2 = \frac{\Omega_2}{2\Delta} \quad (15)$$

Therefore the probability of spontaneous emission loss during the Bragg diffraction process is

$$N_s = \frac{\Omega_2}{2\Delta}\Gamma\tau \quad (16)$$

where $\Omega_0 = \frac{d_{gi}E_0}{\hbar}$ is single-photon Rabi frequency, $d_{gi}$ is electric dipole matrix element connecting atomic ground state ($|g\rangle$) and excited state ($|i\rangle$), $\Omega_2 = \Omega_0^2/2\Delta$ represents the two-photon Rabi frequency, $\Gamma = 2\pi \times 6.06$MHz is the spontaneous decay rate of excited state, and $\tau$ is Bragg pulse duration.

In order to ensure good coherence, the spontaneous emission loss rate should <1%, then the requirement for Bragg pulse frequency detuning is

$$\Delta \geq \frac{\Omega_2 \tau}{2N_s}\Gamma \quad (17)$$

For 1st-order Bragg diffraction, π pulse requires $\Omega_2\tau = \pi$, thus the Bragg pulse frequency detuning should satisfy $\Delta \geq 2\pi \times 1$GHz. For *n*-order Bragg diffraction, π pulse requires $\Omega_{2n}\tau = \pi$, then $\Omega_2\tau \gg \pi$, therefore $\Delta \gg 2\pi \times 1$GHz。

### 3.4 Intensity of Bragg laser

2*n*-photon effective Rabi frequency for *n*-order Bragg diffraction could be deduced on the analogy of the two-photon effective Rabi frequency for 1st-order Bragg diffraction as

$$\Omega_{2n} = \frac{[\Omega_0]^{2n}}{2^{2n-1}\Delta_1\Delta_2\cdots\Delta_{2n-1}} \quad (18)$$

With Eq.(14), taking $\Delta_m \approx \Delta$ when *m* is odd, 2*n*-photon effective Rabi frequency can be simplified to

$$\Omega_{2n} = \frac{[\Omega_2]^n}{(8\omega_r)^{n-1}\left[(n-1)!\right]^2} \quad (19)$$

The relative atom population in state $|p+2n\hbar k\rangle$ after *n*-order Bragg diffraction is

$$P_{|p+2n\hbar k\rangle}(t) = \frac{1}{2}\left[1-\cos(\Omega_{2n}t)\right] \quad (20)$$

Thus, the duration of π and π/2 pulses are

$$\tau_\pi = \pi/\Omega_{2n}, \quad \tau_{\pi/2} = \tau_\pi/2 \quad (21)$$

The duration time $\tau$ of the Bragg laser pulse must satisfy the requirement of Bragg resonance condition together with atomic momentum broadening (see section 3.1), thus should keep within a certain range. Referring to the optimized π pulse duration for *n*-order Bragg diffraction of atoms with transverse momentum of $\Delta p_\perp = \hbar k$ in article [17], the minimum diameter and laser detuning obtained in section 2.2 and 2.3, and parameters of $^{87}$Rb atom, we calculate the optimal two-photon Rabi frequency, intensity and power of the Bragg beam for different diffraction order, as shown in Table 1:

Table 1. Optimal laser parameters of different diffraction order.

| Diffraction order $n$ | | 1 | 5 | 10 | 15 | 20 | 25 |
|---|---|---|---|---|---|---|---|
| Pulse duration $\tau/\omega_r^{-1}$ | | 0.192 | 0.086 | 0.105 | 0.086 | 0.077 | 0.072 |
| Two-photon Rabi frequency $\Omega_2/\omega_r$ | | 16.4 | 38.7 | 118.1 | 254.6 | 443.6 | 685.5 |
| Laser intensity $I$/mW/cm$^2$ | | 18.0 | 42.6 | 130.0 | 280.2 | 488.2 | 754.4 |
| Laser power $P$/mW | BEC atoms | 1.7 | 3.9 | 11.9 | 25.6 | 44.6 | 68.8 |
| | Velocity selective atoms | 5.1 | 12.1 | 36.9 | 79.6 | 138.7 | 214.4 |

From Table 1 we see that the optimized two-photon Rabi frequency, intensity and power of the Bragg beams increase rapidly as the diffraction order $n$ increases. The optimized laser power is much larger for gravimeters using velocity selective atoms than that using BEC atoms. Moreover, experiments are often performed with larger frequency detuning of x GHz (x> 1) in order to reduce the spontaneous emission loss further, thus the optimized intensity and power of the Bragg beams should also be increased by x times. Besides, to reduce the phase error caused by wavefront curvature, the diameter of Bragg laser should be increased further. As a result, the laser power must be increased correspondingly.

### 3.5 Timing of Bragg laser

Because the counter-propagating Bragg SWs usually were produced by mirrors in experiments, as shown in Fig. 5, two pairs of retro-reflected SWs transferring upwards and downwards were formed at the same time. These retro-reflected beams could drive $n$-order Bragg transitions in the opposite direction if atoms are stationary. Letting atoms free fall a time larger than $0.6n$ ms, which assures the atomic Doppler frequency shift becomes greater than the Bragg resonance frequency $n\delta_B$, allows us to select only one of these transitions direction. According to the principles of cold atom interferometry gravimeter based on $n$-order Bragg diffraction described in section 2.3, $\pi/2-\pi-\pi/2$ three Bragg pulses with equal intensity and different duration are applied on atoms separated by time interval $T$ after a free fall time of $t_{\text{Fall}}$. Among each Bragg pulse, one laser is produced at a constant frequency, whilst the other has a frequency difference of $n\delta_B$ at initial time and is triggered to sweep at a rate $\alpha = kg$ to compensate Doppler shift due to gravity. Thus the Bragg beams remain resonant with the atoms as they fall. Combining the conclusions above, timing of intensity and frequency for a Bragg diffraction-based cold atom interferometry gravimeter are shown in Fig. 6.

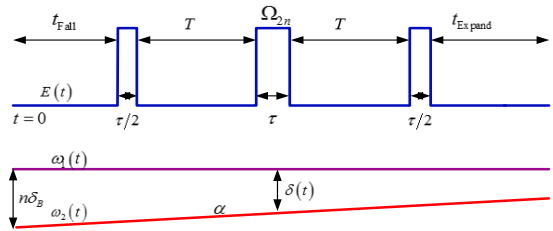

Fig. 6. Timing of intensity and frequency for a Bragg diffraction-based cold atom gravimeter.

## 4 Discussion

Altin et al[12] had presented a precision gravimeter based on coherent Bragg diffraction of freely falling atoms. After Bloch velocity selection atoms with $\Delta p \approx 1\hbar k$, they used Bragg pulses with power 200mW, $1/e^2$ diameter of 7.5 mm, frequency detuning 3GHz, pulse interval $T$=40ms, initial frequency difference $2\delta_B$, frequency chirping rate $\alpha \approx 25.1$MHz/s, achieved a 2nd-order Bragg diffraction atom gravimeter. They obtained a gravity measurement accuracy of $2.7 \times 10^{-9}$g after 1000s integral. Debs et al[13] used Bragg pulses with power 150mW, $1/e^2$ diameter of 3 mm, frequency detuning 90GHz, pulse interval $T$=3ms, initial frequency difference $\delta_B$, frequency chirping rate $\alpha \approx 25.1$MHz/s, achieved a 1st-order Bragg diffraction BEC ($\Delta p \approx 0.14\hbar k$) atom gravimeter. The local gravity they measured is 9.7859(2) m/s$^2$. The above experimental data is consistent with the conclusions of this article.

# 5 Conclusion

By comparing with Raman transition, a new type of cold atom interferometry gravimeter based on *n*-order Bragg diffraction is presented here, which is able to increase the gravity measurement sensitivity by *n* times through LMT. Due to the intrinsic superiority of Bragg diffraction, the common mode error and environmental fluctuations can be diminished effectively. Based on the constructed theoretical model, the key technical requirements have been deduced, which include: longitudinal temperature $T_\parallel \ll 0.36\mu k$; minimum diameter and curvature radius of Bragg laser increase with the transverse temperature of atoms, which should satisfy $w > 6$mm, $R(z) > 48.8$km if a velocity selective atom cloud rather than BEC atom cloud is used; frequency difference of the two Bragg lasers is $\delta_n(t) = n\delta_B + 2kgt$; frequency detuning and intensity of Bragg laser increase rapidly with diffraction order and should satisfy $\Delta \geq 2\pi \times 1$GHz, $I > 18$mW/cm$^2$ when *n*=1; the first Bragg pulse should be applied after atom free fall a certain time to overcome the reverse diffraction. The present study provides a good reference for the understanding and construction of a Bragg atom gravimeter.

## Acknowledgments

This work was supported by the National Natural Science Foundation of China (No. 51275523), the Specialized Research Fund for the Doctoral Program of Higher Education of China (No. 20134307110009), and the Graduate Innovative Research Fund of Hunan Province (No. CX2014A002).